\documentclass[runningheads]{llncs}
\usepackage{amssymb,amsmath}
\usepackage{pstricks}
\usepackage{pst-node}
\usepackage{epsfig}
\usepackage{latexsym}
\begin{document}
\pagestyle{headings}

\newcommand{\vf}[2]{\ensuremath{#1_{#2}}}
\newcommand{\krul}{{\sim}}
\newcommand{\sys}[1]{\ensuremath{\mathbf{#1}}}
\newcommand{\Cn}[2]{\ensuremath{\mathit{Cn}_{\mathbf{#1}}(#2)}}
\newcommand{\Dab}{\ensuremath{\mathit{Dab}}}
\newcommand{\cdic}[1]{#1\wedge\krul #1}
\newcommand{\df}{\ensuremath{=_\mathit{df}}}

\newenvironment{rlist}[1]{%
\begin{list}{}{%
 \settowidth{\labelwidth}{\textrm{#1}}
 \setlength{\labelsep}{1.5\labelsep}  %
 \setlength{\leftmargin}{\labelwidth}
 \addtolength{\leftmargin}{\labelsep}
 \setlength{\parsep}{0cm}
 \setlength{\itemsep}{0cm}
 \setlength{\topsep}{0cm} %
 \renewcommand{\makelabel}[1]{\hfill\textrm{##1}}}}
{\end{list}}

\newenvironment{llist}[1]{%
\begin{list}{}{%
 \settowidth{\labelwidth}{\textrm{#1}}
 \setlength{\labelsep}{1.5\labelsep}    %
 \setlength{\leftmargin}{\labelwidth}
 \addtolength{\leftmargin}{\labelsep}
 \setlength{\parsep}{0cm}
 \setlength{\itemsep}{0cm}
 \renewcommand{\makelabel}[1]{\textrm{##1}\hfill}}}
{\end{list}}

\title{On a Partial Decision Method\\ for Dynamic Proofs%
\thanks{Originally published in proc. PCL 2002, a FLoC workshop;
eds. Hendrik Decker, Dina Goldin, J{\o}rgen Villadsen, Toshiharu Waragai
({\tt http://floc02.diku.dk/PCL/}).}
 \thanks{Research for this paper was supported by
 subventions from Ghent University and from the Fund for Scientific
 Research -- Flanders, and indirectly by the Flemish Minister responsible
 for Science and Technology (contract BIL01/80).
 I am indebted to Dagmar Provijn for comments on a former draft.}
}
 \author{Diderik Batens}
 \institute{Centre for Logic and Philosophy of Science\\
 Universiteit Gent, Belgium\\
 \email{Diderik.Batens@rug.ac.be}}
 \date{}%
\maketitle

\begin{abstract}
This paper concerns a goal directed proof procedure for the propositional fragment of
the adaptive logic \sys{ACLuN1}. At the propositional level, it forms an algorithm for
final derivability. If extended to the predicative level, it provides a \emph{criterion}
for final derivability. This is essential in view of the absence of a positive test. The
procedure may be generalized to all flat adaptive logics.
\end{abstract}

\section{The Problem}  \label{aim}

Inference relations for which there is no positive test abound in both everyday and
scientific reasoning processes. Adaptive logics are intended for characterizing
such inference relations.%
 \footnote{A positive test is a systematic procedure that, for every set of
 premises $\Gamma$ and for every conclusion $A$, leads after finitely many steps to a
 ``yes'' if $A$ is a consequence of $\Gamma$. Remark that the consequence relation
 defined by classical logic is undecidable, but that there is a positive test for it---see
 \cite{B&J89} for such matters.}
The characterization has a specific metalinguistic standard format. This format provides
the logic with a semantics and with a proof theory, and warrants soundness,
completeness, and a set of properties of the logic.%
 \footnote{Only part of these results are written up, viz.\ in \cite{eial}.}
The first adaptive logics were inconsistency-adaptive. The articulation of other
adaptive logics provided increasing insight in the underlying mechanisms and required
that adaptive logics were systematized in a new way. This systematization is presented
in \cite{gcal} and will be followed here.

An especially important feature of adaptive logics is their dynamic proof theory.
Indeed, this proof theory is intended for \emph{explicating} actual reasoning---see
\cite{J:clau1} for a historical example---a task that cannot be accomplished by
definitions, semantic systems, and other more abstract characterizations.

The dynamics of the proof theory provides from the absence of a positive test. For most
consequence relations, the dynamics is double. The \emph{external} dynamics is well
known: as new premises become available, consequences derived from the earlier premise
set may be withdrawn. In other words, the external dynamics results from the
non-monotonic character of the consequence relation---the fact that, for some $\Gamma$,
$\Delta$ and $A$, $\Gamma \vdash A$ but $\Gamma \cup \Delta \not\vdash A$. The
\emph{internal} dynamics is very different from the external one. Even if the premise
set is constant, certain formulas are considered as derived at some stage of the proof,
but are considered as not derived at a later stage. For any consequence relation,
insight in the premises is only gained by deriving consequences from them. In the
absence of a positive test, this results in the internal dynamics.%
 \footnote{The Weak consequence relation from \cite{NR:hr} and \cite{R&M}---see
 \cite{BD&P1} and \cite{BD&P2} for an extensive study of such consequence relations---is
 monotonic. Nevertheless, its proof theory necessarily displays an internal dynamics
 because there is no positive test for it---see \cite{unific} and \cite{RMs}. Some logics
 for which there is a positive test, may nevertheless be characterized in a nice way in
 terms of a dynamic proof theory---see \cite{Rpijl}.}

Dynamic proofs differ in two main respects from usual proofs. The first difference
concerns annotated versions. Apart from (i)~a line number, (ii)~a formula, (iii)~the
line numbers of the formulas from which the formula is derived, and (iv)~the rule by
which the formula is derived (the latter two are the justification of the line), dynamic
proofs also contain (v)~a \emph{condition}. Intuitively, this is a set of formulas that
are supposed to be false, or, to be more precise, formulas the truth of which is not
required by the premises.

The second main difference is that, apart from the deduction rules that allow one to add
lines to the proof, there is a \emph{marking definition}. The underlying idea is as
follows. As the proof proceeds, more formulas are derived from the premises. In view of
these formulas, some conditions may turn out not to hold. The lines at which such
conditions occur are \emph{marked}. Formulas derived on marked lines are taken not to be
derived from the premises. In other words, they are considered as `out'. One way to
understand the procedure is as follows. As the proof proceeds, one's insight in the
premises improves. More particularly, some of the conditions that were introduced
earlier may turn out not to hold.

For any stage of the proof, the marking definition settles which lines are marked and
which lines are unmarked. This leads to a precise definition of \emph{derivability at a
stage}. Notwithstanding the precise character of this notion, we also want a more stable
form of derivability, which is called \emph{final derivability}. The latter does not
depend on the stage of the proof; nor does it depend on the way in which a specific
proof from a set of premises proceeds. It is an abstract and stable relation between a
set of premises and a conclusion. A different way for putting this is that final
derivability refers to a stage of the proof at which the marks have become stable. Final
derivability should be sound and strongly complete with respect to the semantics. For
any adaptive logic \sys{AL}, $A$ should be finally derivable from $\Gamma$ ($\Gamma
\vdash_\sys{AL} A$) if and only if $A$ is a semantic consequence of $\Gamma$ ($\Gamma
\vDash_\sys{AL} A$).

Consider a dynamic proof from a set of premises. At any point in time, the proof will be
finite. It will reveal what is derivable from the premises at that stage of the proof.
But obviously we are interested in final derivability. Whence the question: what does a
proof at a stage reveal about final derivability? As there is no positive test for the
consequence relation, there is no algorithm for final derivability. So, one has at best
some \emph{criteria} that decide, for specific $A$ and $\Gamma$, whether $A$ is finally
derivable from $\Gamma$.

What if no criterion enables one to conclude from the proof whether certain formulas are
or are not finally derivable from the premise set? The answer or rather the answers to
this question are presented in \cite{blocks}. Roughly, they go as follows. First, there
is a characteristic semantics for derivability at a stage. Next, it can be shown that,
as the dynamic proof proceeds, the insight in the premises provided by the proof
never decreases and may increase.%
 \footnote{More particularly, this insight increases if informative steps are added to
 the proof, where ``informative step'' is clearly definable---see \cite{blocks}.}
In other words, derivability at a stage provides an estimate for final derivability,
and, as the proof proceeds, this estimate may become better, and never becomes worse. In
view of all this, derivability at a stage gives one exactly what one might expect, viz.\
a fallible but \emph{sensible} estimate of final derivability.%
 \footnote{This estimate is defined in terms of the proof theory, and the latter
 explicates actual reasoning. So, the estimate should not be confused with
 approximations that may be obtained by certain computational procedures.}
At any stage of the proof, one has to decide (obviously on the basis of pragmatic
considerations) whether one will continue the proof or rely on present insights.
This is fully in line with the contemporary view on rationality.%
 \footnote{Needless to say, some proofs provide more efficient estimates of final
 derivability than others. The goal directed proofs presented in this paper offer means
 to obtain efficient proofs, but more research on this problem is desirable.}

Needless to say, one should apply a criterion for final derivability whenever one can.
This motivated the search for such criteria---see \cite{blocks}, \cite{metJ:tabl1} and
\cite{metJ:tabl2}. Unfortunately, most of these criteria are complex and only
transparent for people that are well acquainted with dynamic proofs. Recently, it turned
out that a specific kind of goal directed proofs offer a way out in this respect. The
idea is not to formulate a criterion, but rather to specify a specific proof procedure
that functions as a criterion. The proof procedure is applied to $\Gamma \vdash_\sys{AL}
A$. Whenever the proof procedure stops, it is possible to conclude from the resulting
proof whether or not $\Gamma \vdash_\sys{AL} A$. Preparatory work on the propositional
fragment of \sys{CL} (classical logic) is presented in \cite{metD:vd1} and some first
results on the proof procedure for inconsistency-adaptive logics are presented in this
paper.

The present paper is restricted to the propositional level. So, all references to
logical systems concern the propositional fragments only. At this level the proof
procedure forms an \emph{algorithm} for final derivability: if the proof procedure is
applied to $A_1, \ldots , A_n \vdash B$, it always stops after finitely many steps. If,
at the last stage of the proof, $B$ is derived on an unmarked line, then $B$ is finally
derivable from $A_1, \ldots , A_n$; if $B$ is not derived on an unmarked line, it is not
finally derivable from $A_1, \ldots , A_n$. However, the proof procedure may be extended
to the predicative level and there provides a criterion for final derivability if it
stops. The main interest of the procedure lies there.

In Section \ref{aclun1}, I briefly present the inconsistency-adaptive logic \sys{ACLuN1}
and its dynamic proof theory. In Section \ref{gdp}, the goal-directed proof is applied
to \sys{CL}. This will make the matter easily understood by everyone. The proof
procedure for the adaptive logic \sys{ACLuN1} is spelled out in Section \ref{gdpaclun}.

\section{The Inconsistency-Adaptive Logic \sys{ACLuN1}}  \label{aclun1}

The central difference between paraconsistent logics and inconsistency-adaptive logics
can be easily described in proof theoretic terms. In a (monotonic) paraconsistent logic
some deduction \emph{rules} of \sys{CL} are invalid; in an inconsistency-adaptive logic,
some \emph{applications} of deduction rules of \sys{CL} are invalid.

The original application context that led to inconsistency-adaptive logics---see
\cite{ddl}---is still one of the most clarifying ones. Suppose that a theory $T$ was
intended as consistent and was formulated with \sys{CL} as its underlying logic. Suppose
next that $T$ turns out to be inconsistent. Of course, one will want to replace $T$ by
some consistent improvement $T'$. Typically, one does not just trow away $T$, restarting
from scratch. One \emph{reasons} from $T$ in order to locate the inconsistency or
inconsistencies and in order to locate constraints for the replacement $T'$. Needless to
say, logic alone is not sufficient to find the justified replacement $T'$.%
 \footnote{If $T$ is an empirical theory, at least new factual data (observations and
 outcomes of experiments) will be required. If $T$ is a mathematical theory, more
 conceptual analysis will be required.}
However, logic is able to \emph{locate} the inconsistencies in $T$. It can provide one
with an interpretation of $T$ that is `as consistently as possible'. Let me phrase this
in intuitive terms. At points where $T$ is inconsistent, some deduction rules of
\sys{CL} cannot apply---if they did, the resulting interpretation of $T$ would be
trivial in that it would make \emph{every} sentence of the language a theorem of $T$.
But where $T$ is consistent, all deduction rules of \sys{CL} should apply.

An extremely simple propositional example will clarify the matter. Consider the theory
$T$ that is characterized by the premise set $\{p, \krul p\vee r, q, \krul q\vee s,
\krul p\}$. From these premises, $r$ should not be derived by Disjunctive Syllogism.
Indeed, $\krul p\vee r$ is just an obvious weakening of $\krul p$. If one were to derive
$r$ from the premises, then, by the same reasoning, one should derive $\krul r$ from $p$
and $\krul p\vee \krul r$, which also is an obvious weakening of $\krul p$. However, if
one interprets the premises as consistently as possible, one should derive $s$ from
them, viz.\ by Disjunctive Syllogism from $q$ and $\krul q \vee s$. Indeed, while the
premises require $p$ to behave inconsistently (require $p\wedge \krul p$ to be true),
they do not require $q$ to behave inconsistently (they do not require $q\wedge \krul q$
to be true).

As the matter is central, let me phrase it differently. The theory $T$ from the previous
paragraph turns out to be inconsistent. As it was intended to be consistent, it should
be interpreted as consistently as possible. Given that $T$ is inconsistent, one will
move `down' to a paraconsistent logic---a logic that allows for inconsistencies. If a
formula turns out to be inconsistent on the paraconsistent reading of $T$, one cannot
apply certain rules of \sys{CL} to it. Thus, even on the paraconsistent interpretation
of $T$, $p\wedge \krul p$ is true. But consider $p\wedge (\krul p\vee r)$. Given the
meaning of conjunction and disjunction, this formula is equivalent to $(p\wedge \krul
p)\vee r$. According to \sys{CL}, $p\wedge \krul p$ cannot be true, and hence $r$ is
true. However, the premises state that $p\wedge \krul p$ is true. So, if one wants to
reason sensibly \emph{from} these premises, one cannot rely on the
\sys{CL}-presupposition that $p\wedge \krul p$ is bound to be false. However, where the
paraconsistent reading of $T$ does not require that a specific formula $A$ behaves
inconsistently, one may retain the \sys{CL}-presupposition that $A$ is consistent, and
hence apply \sys{CL}-rules where they are validated by \emph{this} presupposition. Thus
$T$ affirms $q\wedge (\krul q\vee s)$, which is equivalent to $(q\wedge \krul q)\vee s$.
As $T$ does not require $q\wedge \krul q$ to be true, it should be taken to be false and
one should conclude to $s$.

The intuitive statements from the two preceding paragraphs are given a precise and
coherent formulation by inconsistency-adaptive logics.

An adaptive logic is characterized by the following triple:%
 \footnote{In this paper I consider only \emph{flat} adaptive logics.
 Other adaptive logics are the prioritized ones, which are defined as specific
 combinations of flat adaptive logics---see \cite{gcal}.}%
\begin{rlist}{(iiii)}
\item[(i)]
 a monotonic \emph{lower limit logic},
\item[(ii)]
 a \emph{set of abnormalities} (characterized by a logical form), and
\item[(iii)]
 an \emph{adaptive strategy} (specifying the meaning of ``interpreting the premises
 as normally as possible'').
\end{rlist}
Extending the lower limit logic with the requirement that no abnormality is logically
possible results in a monotonic logic, which is called the \emph{upper limit logic}.

Let me illustrate this by the specific inconsistency-adaptive logic \sys{ACLuN1}. In
this paper, I shall only consider the propositional level of the logic and I shall
consider no other strategy than Reliability.

The \emph{lower limit logic} of \sys{ACLuN1} is \sys{CLuN}. This monotonic
paraconsistent logic is just like \sys{CL}, except in that it allows for gluts with
respect to negation---whence the name \sys{CLuN}. Axiomatically, \sys{CLuN} is obtained
by extending full positive propositional logic with the axiom schema $A\vee\krul
A$---see \cite{ial} for a study of the full logics \sys{CLuN} and \sys{ACLuN1},
including the semantics. \sys{CLuN} isolates inconsistencies. Indeed, Double Negation,
de Morgan rules, and all similar negation reducing rules are \emph{not} validated by
\sys{CLuN}. As a result, complex contradictions do not reduce to truth functions of
simpler contradictions.%
 \footnote{For example, $\cdic{(p\wedge q)} \not\vdash_\sys{CLuN} (\cdic{p})\vee(\cdic{q})$
 and $\cdic{\krul p} \not\vdash_\sys{CLuN} \cdic{p}$. Of course, one still has
 $\cdic{(\cdic{p})} \vdash_\sys{CLuN} \cdic{p}$.}
There are several versions of \sys{CLuN}. I shall suppose that the language contains
$\bot$, characterized by the axiom schema $\bot \supset A$, and I shall discuss this
convention below.

The \emph{set of abnormalities}, $\Omega$, comprises all formulas%
 \footnote{For some logics, the abnormalities are couples consisting of an
 open formula with $n$ free variables and of an $n$-tuple of elements of the domain.}
of the form $\cdic{A}$.%
 \footnote{Some flat adaptive logics are described and studied as formula-preferential
 systems in \cite{IL:thesis}---see also---\cite{A&L:fps}. $\Omega$ is then any set of
 formulas. It is not clear whether this may be generalized to all adaptive logics.}
Extending \sys{CLuN} with the axiom schema $(\cdic{A})\supset B$ results in the
\emph{upper limit logic}, which is \sys{CL}.

Finally, we come to the adaptive \emph{strategy}. Below I shall often need to refer to
\emph{disjunctions of abnormalities}, which I shall call \Dab-formulas. From now on an
expression of the form $\Dab(\Delta)$ will refer to a disjunction of abnormalities; in
other words, $\Delta$ is a finite subset of $\Omega$
and $\Dab(\Delta)$ is the disjunction of the members of $\Delta$.%
 \footnote{It can be shown that $\Gamma \vdash_\sys{CL} \bot$ iff there is a finite
 $\Delta \subset \Omega$ such that $\Gamma \vdash_\sys{CLuN} \Dab(\Delta)$. So, both
 expressions may be taken to define that $\Gamma$ is inconsistent.}
Suppose now that $\Gamma \vdash_\sys{CLuN} \Dab(\Delta)$, but that no member of $\Delta$
is \sys{CLuN}-derivable from $\Gamma$. This means that the premises require some member
of $\Delta$ to be true, but do not specify which member is true. In view of this
possibility, one needs to introduce an adaptive strategy. One wants to interpret the
premises ``as normally as possible'' (which for the present $\Omega$ means ``as
consistently as possible''), but this phrase is ambiguous. As indicated in (iii), an
adaptive strategy disambiguates the phrase.

The \emph{Reliability strategy} from \cite{ddl}%
 \footnote{This is the oldest paper on adaptive logics, but it appeared in a book that
 took ten years to come out.}
is the oldest known strategy, and the one that is simplest from a proof theoretic point
of view. I shall not consider any other strategies in this paper. Let $\Dab(\Delta)$ be
a \emph{minimal \Dab-consequence of} $\Gamma$ iff $\Gamma \vdash_\sys{CLuN}
\Dab(\Delta)$ and there is no $\Delta' \subset \Delta$ for which $\Gamma
\vdash_\sys{CLuN} \Dab(\Delta')$. Let $U(\Gamma) = \{ A \mid A \in \Delta$ for some
minimal \Dab-consequence $\Dab(\Delta)$ of $\Gamma \}$ be the set of formulas that are
\emph{unreliable} with respect to $\Gamma$. Below, I shall define $\Gamma
\vdash_\sys{ACLuN1} A$, which will be read as ``$A$ is finally \sys{ACLuN1}-derivable
from $\Gamma$''. The following Theorem is provable. In plain words it says that $A$ is
\sys{ACLuN1}-derivable from $\Gamma$ iff there is a $\Delta$ such that $A\vee
\Dab(\Delta)$ is \sys{CLuN}-derivable from $\Gamma$ and no member of $\Delta$ is
unreliable with respect to $\Gamma$.

\begin{theorem}
 $\Gamma \vdash_\sys{ACLuN1} A$
 iff
 there is a $\Delta \subseteq \Omega$ such that
 $\Gamma \vdash_\sys{CLuN} A\vee \Dab(\Delta)$
 and
 $\Delta \cap U(\Gamma) = \emptyset$.
\end{theorem}

The dynamic proof theory of any (flat) adaptive logic is characterized by three
(generic) rules, except of course that the rules RU and RC should refer to the right
lower limit logic. Let $\Gamma$ be the set of premises as before. I now list the
official deduction rules.%
 \footnote{Only RC introduces non-empty conditions. In other words, as long as RC is not
 applied, the condition of every line is $\emptyset$.}
Immediately thereafter I shall mention a shorthand notation that most people will find
more transparent.

\begin{llist}{PREM}
\item[PREM]
 If $A \in \Gamma$, one may add a line comprising the following elements:
 (i)~an appropriate line number,
 (ii)~$A$,
 (iii)~$-$,
 (iv)~PREM, and
 (v)~$\emptyset$.
\item[RU]
 If $A_1, \ldots, A_n \vdash_\sys{CLuN} B$
 and each of $A_1$, $\ldots$, $A_n$ occur in the proof
 on lines $i_1$, \ldots, $i_n$
 that have conditions $\Delta_1$, $\ldots$, $\Delta_n$ respectively,
 one may add a line comprising the following elements:
 (i)~an appropriate line number,
 (ii)~$B$,
 (iii)~$i_1, \ldots, i_n$,
 (iv)~RU, and
 (v)~$\Delta_1 \cup \ldots \cup \Delta_n$.
\item[RC]
 If $A_1, \ldots, A_n \vdash_\sys{CLuN} B\vee \Dab(\Theta)$
 and each of $A_1$, $\ldots$, $A_n$ occur in the proof
 on lines $i_1$, \ldots, $i_n$
 that have conditions $\Delta_1$, $\ldots$, $\Delta_n$ respectively,
 one may add a line comprising the following elements:
 (i)~an appropriate line number,
 (ii)~$B$,
 (iii)~$i_1, \ldots, i_n$,
 (iv)~RC, and
 (v)~$\Delta_1 \cup \ldots \cup \Delta_n \cup \Theta$.
\end{llist}

Where ``$A$ \quad $\Delta$'' abbreviates that $A$ occurs in the proof on the condition
$\Delta$, the rules may be phrased more transparently as follows:
\bigskip

\noindent
\begin{tabular}[t]{lll}
PREM~~ & If $A \in \Gamma$: &
 \begin{tabular}[t]{ll}
  \ldots & \ldots \\ \hline
  $A$ & $\emptyset$
 \end{tabular}\\
 \\
RU & If $A_1, \ldots, A_n \vdash_\sys{LLL} B$: &
 \begin{tabular}[t]{ll}
  $A_1$ & $\Delta_1$ \\
   \ldots & \ldots \\
  $A_n$ & $\Delta_n$ \\ \hline
  $B$ & $\Delta_1 \cup \ldots \cup \Delta_n$
 \end{tabular}\\
 \\
RC & If $A_1, \ldots, A_n \vdash_\sys{LLL} B\vee \Dab(\Theta)$:~~ &
 \begin{tabular}[t]{ll}
  $A_1$ & $\Delta_1$ \\
   \ldots & \ldots \\
  $A_n$ & $\Delta_n$ \\ \hline
  $B$ & $\Delta_1 \cup \ldots \cup \Delta_n \cup \Theta$
 \end{tabular}
\end{tabular}
\bigskip

While the deduction rules enable one to add lines to the proof, the marking definition,
which depends on the strategy, determines which lines are ``in'' and which lines are
``out''. For the Reliability strategy, we first need to define the set $U_s(\Gamma)$ of
formulas that are unreliable at a stage $s$ of a proof. Let $\Dab(\Delta)$ be a
\emph{minimal \Dab-formula} at stage $s$ of the proof iff, at that stage, $\Dab(\Delta)$
has been derived on the condition $\emptyset$ \emph{and} there is no $\Delta' \subset
\Delta$ for which $\Dab(\Delta')$ has been derived on the condition $\emptyset$.%
 \footnote{The minimal \Dab-formulas that occur in a proof at a stage should not be confused
 with minimal \Dab-consequences of the set of premises. At a stage $s$, a new minimal
 \Dab-formula may be derived, and the effect may be that a \Dab-formula that was minimal
 at stage $s-1$ is not minimal at stage $s$. Whether some formula is a minimal
 \Dab-consequence of the premises is obviously independent of the stage of a proof from
 those premises.}
Let $U_s(\Gamma) \df \{ A \mid A \in \Delta$ for some minimal \Dab-formula
$\Dab(\Delta)$ at stage $s$ of the proof$\,\}$.

\begin{definition} \label{d:mrel}
 Where $\Delta$ is the condition of line $i$,
 line $i$ is marked at stage $s$
 iff
 $\Delta \cap U_s(\Gamma) \neq \emptyset$.
 \quad
 (Marking definition for Reliability)
\end{definition}

Lines that are unmarked at one stage may be marked at the next, and vice versa. Finally,
I list the definitions that concern final derivability---the definitions are identical
for all adaptive logics.

\begin{definition} \label{d:fdas}
 $A$ is \emph{finally derived} from $\Gamma$ on line $i$ of a proof at stage $s$
 iff
 $A$ is derived on line $i$,
 line $i$ is not marked at stage $s$, and
 any extension of the proof in which line $i$ is marked
 may be further extended in such a way that line $i$ is unmarked.
\end{definition}

\begin{definition} \label{d:fd}
 $\Gamma \vdash_\sys{AL} A$ ($A$ is \emph{finally} \sys{AL}-\emph{derivable} from $\Gamma$)
 iff
 $A$ is finally derived on a line of a proof from $\Gamma$.
\end{definition}

Remark that by ``a proof'' I mean (here and elsewhere) a sequence of lines that is
obtained by applying certain instructions. In the present context, this means that each
line in the sequence is obtained by applying a deduction rule and that the marking
definition was applied. Here is a very simple dynamic proof.

\begin{tabbing}
00Mm \= $(\cdic{p})\vee(\cdic{p}) $m \= 00, 00 \= Premmm \= \kill
 1 \> $(p\wedge q)\wedge t$ \> $-$ \> PREM \> $\emptyset$\\
 2 \> $\krul p\vee r$ \> $-$ \> PREM \> $\emptyset$\\
 3 \> $\krul q\vee s$ \> $-$ \> PREM \> $\emptyset$\\
 4 \> $\krul p\vee \krul q$ \> $-$ \> PREM \> $\emptyset$\\
 5 \> $t\supset \krul p$ \> $-$ \> PREM \> $\emptyset$\\
 6 \> $r$ \> 1, 2 \> RC \> $\{\cdic{p}\}^{\fbox{\tiny 9}}$\\
 7 \> $s$ \> 1, 3 \> RC \> $\{\cdic{q}\}$\\
 8 \> $(\cdic{p})\vee(\cdic{q})$ \> 1, 4 \> RU \> $\emptyset$\\
 9 \> $\cdic{p}$ \> 1, 5 \> RU \> $\emptyset$
\end{tabbing}

Up to stage 7 of the proof, all lines are unmarked. At stage 8, lines 6 and 7 are marked
because $U_8(\Gamma) = \{\cdic{p}, \cdic{q}\}$. At stage 9, only line 6 is marked
because $U_9(\Gamma) = \{\cdic{p}\}$. It is easily seen that, if 1--5 are the only
premises, then the marks will remain unchanged in all extensions of the proof. So, $r$
is not a final consequence of $\Gamma$ whereas $s$ is a final consequence of $\Gamma$.

\paragraph{The convention on $\bot$.}
As promised, I now discuss the convention that the language contains $\bot$ and hence
that classical negation can be defined within the language, viz.\ by $\neg A \df
A\supset \bot$. In a sense then, \sys{CLuN} is an extension of \sys{CL}. It has the full
inferential power of \sys{CL}, $\neg$ functioning as the \sys{CL}-negation, and moreover
contains the paraconsistent negation $\krul$. In the original application context,
mentioned in the second paragraph of this section, the premises belong to the
$\bot$-free and $\neg$-free fragment of the language---of course, there are different
application contexts as well. However, even in the original application context the
presence of $\neg$ is useful: it greatly simplifies metatheoretic proofs and technical
matters in general, and in no way hampers the limitations imposed by the application
context.%
 \footnote{By present lights, it is harmless as well as useful, for all adaptive logics,
 to extend the language and the lower limit logic in such a way that all classical
 connectives belong to the lower limit logic. This holds even if these connectives
 do not occur in the premises or in the conclusions a user is interested in---see
 \cite{CL0} for an example.}
As will appear in Section \ref{gdpaclun}, the presence of $\neg$ also greatly simplifies
the goal directed proof procedure that will serve as a criterion for final derivability.

\section{Goal Directed Proofs for Classical Logic} \label{gdp}

In this section I merely present an example: a goal directed proof for $p\supset
(q\wedge s), \neg (q\vee r) \vdash_\sys{CL} \neg p$.%
 \footnote{In order to avoid useless complications, I write classical negation
 as $\neg$ even in the context of \sys{CL}.}
As the proof is simple, I skip the rules as well as the heuristic instructions---these
are spelled out in \cite{metD:vd1}---and merely offer some comments.

The first step introduces the main goal:
\begin{tabbing}
00Mm \= $(\cdic{p})\vee(\cdic{p}) $m \= 00, 00 \= Premmm \= \kill
 1 \> $\neg p$ \>  \> GOAL \> $\{\neg p\}$
\end{tabbing}
This step is meant to remind one that one is looking for the formula that occurs in the
condition, viz.\ $\neg p$. Remark that the purpose served by a condition in goal
directed proofs is very different from the one in dynamic proofs---it is `prospective'
rather than `defeasible'. In view of the condition of line 1, one introduces a premise
from which $\neg p$ may be obtained, and next analyses the premise:
\begin{tabbing}
00Mm \= $(\cdic{p})\vee(\cdic{p}) $m \= 00, 00 \= Premmm \= \kill
 2 \> $p\supset (q\wedge s)$ \>  \> PREM \> $\emptyset$\\
 3 \> $\neg p$ \> 2 \> $\supset$E \> $\{\neg (q\wedge s)\}$
\end{tabbing}
Line 3 illustrates a formula analysing rule: in view of 2, one would have $\neg p$ if
one had $\neg (q\wedge s)$. As $\neg (q\wedge s)$ cannot be obtained by analysing a
premise, one applies a condition analysing rule to $\neg (q\wedge s)$:
\begin{tabbing}
00Mm \= $(\cdic{p})\vee(\cdic{p}) $m \= 00, 00 \= Premmm \= \kill
 4 \> $\neg p$ \> 3 \> C$\neg\wedge$E \> $\{\neg q\}$
\end{tabbing}
The following steps require no comment:
\begin{tabbing}
00Mm \= $(\cdic{p})\vee(\cdic{p}) $m \= 00, 00 \= Premmm \= \kill
 5 \> $\neg (q\vee r)$ \>  \> PREM \> $\emptyset$\\
 6 \> $\neg q$ \> 5 \> $\neg\vee$E \> $\emptyset$\\
 7 \> $\neg p$ \> 4, 6 \> Trans \> $\emptyset$
\end{tabbing}
As the main goal is obtained on the empty condition at line 7, the proof is completed.

It is easily seen that, in a proof for $\Gamma \vdash_\sys{CL} A$, a formula $B$ is
derivable on the condition $\Delta$ just in case $\Gamma \cup \Delta \vdash_\sys{CL} B$.

Some lines are marked in goal-directed proofs for \sys{CL}. Unlike what was the case in
the previous section, these marks indicate that one should not try to derive the members
of the condition of marked lines. More details are presented in the next section, where
these marks will be called D-marks because they relate to derivability---A-marks will
relate to the adaptive character of the logic.

\section{Goal Directed Proofs for \sys{ACLuN1}} \label{gdpaclun}

Goal directed proofs for \sys{ACLuN1} have lines that contain \emph{two} conditions:
\begin{tabbing}
00Mm \= $(\cdic{p})\vee(\cdic{p}) $m \= 00, 00 \= Premmm \= $\Delta$MMM \=\kill
 $i$ \> $A$ \> \ldots \> \ldots \> $\Delta$ \> $\Theta$
\end{tabbing}
The first, $\Delta$, is called the D-condition. This is the condition that also occurs
in goal directed proofs for \sys{CL}; it contains the formulas that one needs to derive
in order to obtain $A$. The second condition, $\Theta$, is called the A-condition; it
contains the abnormalities that should not belong to $U(\Gamma)$ in order for $A$ to be
derivable from the premises. The occurrence of the above line $i$ in a proof from
$\Gamma$ warrants that $\Gamma \cup \Delta \vdash_\sys{CLuN} A\vee \Dab(\Theta)$. In
order to show that $\Gamma \vdash_\sys{ACLuN1} G$ one needs a line like the displayed
one at which $A = G$, $\Delta = \emptyset$, and $\Theta \cap U(\Gamma) = \emptyset$.

To facilitate the exposition, I shall write $A_{\Delta,\Theta}$ to denote that $A$ has
been derived on the D-condition $\Delta$ and on the A-condition $\Theta$.

Let us first consider the plot. A goal directed \sys{ACLuN1}-proof for $A_1, \ldots
,\linebreak[1] A_n \vdash G$ will consist of three phases. In the first phase, one tries
to obtain $G_{\emptyset,\Theta}$ for some $\Theta$---this phase starts by an application
of the Goal rule. If this succeeds, one moves on to phase 2 and tries to obtain
$\Dab(\Theta)_{\emptyset,\Lambda}$ for some $\Lambda$---this phase starts by an
application of the A-Goal rule. If this succeeds, one moves on to phase 3 and tries to
obtain $\Dab(\Lambda)_{\emptyset,\emptyset}$---this phase starts by an application of
the X-Goal rule. If, in phase 3, the X-Goal is reached or the procedure stops, one
returns to phase 2; if the procedure stops in phase 2, one returns to phase 1. In phase
1, there are two subphases: 1A and 1B; subphase 1B is introduced by the first
application of EFQ. Details are given below.

Four kinds of rules govern a proof for $\Gamma \vdash_\sys{ACLuN1} G$. The following
rules introduce premises or start new phases or subphases of the proof. A-Goal and
X-Goal are identical but are used in different contexts.
\begin{llist}{TransMM}  \setlength{\itemsep}{1ex}
\item[Prem]
 If $A\in \Gamma$, introduce $A_{\emptyset,\emptyset}$.
\item[Goal]
 Introduce $G_{\{G\},\emptyset}$.
\item[A-Goal]
 If $\Delta \subseteq \Omega$, introduce $\vf{\Dab(\Delta)}{\{\Dab(\Delta)\}, \emptyset}$.
\item[X-Goal]
 If $\Delta \subseteq \Omega$, introduce $\vf{\Dab(\Delta)}{\{\Dab(\Delta)\}, \emptyset}$.
\item[EFQ]
 If $A\in \Gamma$, introduce $G_{\{\neg A\},\emptyset}$.
\end{llist}\medskip

Formula analysing rules (two formulas below the line indicate variants):\bigskip

\noindent
\begin{tabular}{@{}llll}
   $\supset$E~
 & \begin{tabular}{c}
    $\vf{(A \supset B)}{\Delta,\Theta}$\\ \hline
    $\vf{B}{\Delta \cup \{A\},\Theta}\quad\vf{\neg A}{\Delta \cup \{\neg B\},\Theta}$
   \end{tabular}
 & $\neg$$\supset$E~
 & \begin{tabular}{l}
    $\vf {\neg (A \supset B)} {\Delta,\Theta}$\\ \hline
    $\vf {A} {\Delta,\Theta} \quad \vf {\neg B} {\Delta,\Theta}$
   \end{tabular}
 \medskip \\
   $\vee$E
 & \begin{tabular}{c}
    $\vf{(A \vee B)}{\Delta,\Theta}$\\ \hline
    $\vf{A}{\Delta \cup \{\neg B\},\Theta}\quad\vf{B}{\Delta \cup \{\neg A\},\Theta}$
   \end{tabular}
 & $\neg\vee$E
 & \begin{tabular}{l}
    $\vf {\neg (A \vee B)} {\Delta,\Theta}$\\ \hline
    $\vf {\neg A} {\Delta,\Theta} \quad \vf {\neg B} {\Delta,\Theta}$
   \end{tabular}
 \medskip \\
   $\wedge$E
 & \begin{tabular}{c}
    $\vf {(A \wedge B)} {\Delta,\Theta}$\\ \hline
    $\vf {A} {\Delta,\Theta} \quad \vf {B} {\Delta,\Theta}$
   \end{tabular}
 & $\neg\wedge$E
 & \begin{tabular}{c}
    $\vf {\neg (A \wedge B)} {\Delta,\Theta}$\\ \hline
    $\vf {(\neg A \vee \neg B)} {\Delta,\Theta}$
   \end{tabular}
 \medskip \\
   $\equiv$E
 & \begin{tabular}{c}
    $\vf {(A \equiv B)} {\Delta,\Theta}$\\ \hline
    $\vf {(A \supset B)} {\Delta,\Theta} \quad \vf {(B \supset A)} {\Delta,\Theta}$
   \end{tabular}~~~~~~~~
 & $\neg$$\equiv$E
 & \begin{tabular}{c}
    $\vf {\neg (A \equiv B)} {\Delta,\Theta}$\\ \hline
    $\vf {(A \vee B)} {\Delta,\Theta} \quad \vf {(\neg A \vee \neg B)} {\Delta,\Theta} $
   \end{tabular}
 \medskip \\
   $\krul$E
 & \begin{tabular}{c}
    $\vf{\krul A}{\Delta, \Theta}$ \\ \hline
    $\vf{\neg A}{\Delta, \Theta\cup \{\cdic{A}\}}$
   \end{tabular}
 & $\neg\krul$E
 & \begin{tabular}{c}
    $\vf{\neg\krul A }{\Delta,\Theta}$\\ \hline
    $\vf{A}{\Delta,\Theta}$
   \end{tabular}
 \medskip \\
 && $\neg\neg$E
 & \begin{tabular}{c}
    $\vf {\neg\neg A} {\Delta,\Theta}$\\ \hline
    $\vf {A} {\Delta,\Theta}$
   \end{tabular}
\end{tabular}
\bigskip

Condition analysing rules:\bigskip

\noindent
\begin{tabular}{@{}llll}
   C$\supset$E~
 & \begin{tabular}{c}
    $\vf{A} {\Delta \cup \{B \supset C\},\Theta}$\\ \hline
    $\vf{A}{\Delta \cup \{\neg B\},\Theta}\quad \vf{A}{\Delta \cup \{C\},\Theta}$
   \end{tabular}~~~~~~~~~~
 & C$\neg$$\supset$E~
 & \begin{tabular}{c}
    $\vf{A}{\Delta \cup \{\neg (B \supset C)\},\Theta}$\\ \hline
    $\vf{A}{\Delta \cup \{B, \neg C\},\Theta}$
   \end{tabular}
\end{tabular}

\noindent
\begin{tabular}{@{}llll}
   C$\vee$E
 & \begin{tabular}{c}
    $\vf{A}{\Delta \cup \{B \vee C\},\Theta}$\\ \hline
    $\vf{A}{\Delta \cup \{B\},\Theta} \quad \vf{A}{\Delta \cup \{C\},\Theta}$
   \end{tabular}
 & C$\neg\vee$E
 & \begin{tabular}{c}
    $\vf{A}{\Delta \cup \{\neg (B \vee C)\},\Theta}$\\ \hline
    $\vf{A}{\Delta \cup \{\neg B, \neg C\},\Theta}$
   \end{tabular}
 \medskip \\
   C$\wedge$E
 & \begin{tabular}{c}
    $\vf{A}{\Delta \cup \{B \wedge C\},\Theta}$\\ \hline
    $\vf{A}{\Delta \cup \{B,C\},\Theta}$
   \end{tabular}
 & C$\neg\wedge$E
 & \begin{tabular}{c}
    $\vf{A}{\Delta \cup \{\neg (B \wedge C)\},\Theta}$\\ \hline
    $\vf{A}{\Delta \cup \{\neg B\},\Theta}\quad \vf{A}{\Delta \cup \{\neg C\},\Theta}$
   \end{tabular}
 \medskip \\
   C$\equiv$E
 & \begin{tabular}{c}
    $\vf{A} {\Delta \cup \{B \equiv C\},\Theta}$\\ \hline
    $\vf{A}{\Delta \cup \{B,C\},\Theta}\quad \vf{A}{\Delta \cup \{\neg B, \neg C\},\Theta}$
   \end{tabular}~~~~
 & C$\neg$$\equiv$E
 & \begin{tabular}{c}
    $\vf{A}{\Delta \cup \{\neg (B \equiv C)\},\Theta}$\\ \hline
    $\vf{A}{\Delta \cup \{\neg B, C\},\Theta}\quad \vf{A}{\Delta \cup \{B, \neg C\},\Theta}$
   \end{tabular}
 \medskip \\
   C$\krul$E
 & \begin{tabular}{c}
    $\vf{A}{\Delta \cup \{\krul B\},\Theta}$ \\ \hline
    $\vf{A}{\Delta \cup \{\neg B\},\Theta}$
   \end{tabular}
 & C$\neg\krul$E
 & \begin{tabular}{c}
    $\vf{A}{\Delta\cup \{\neg\krul B\}, \Theta }$ \\ \hline
    $\vf{A}{\Delta \cup \{ B\}, \Theta\cup \{\cdic{B}\}}$
   \end{tabular}
\medskip \\
 && C$\neg\neg$E
 & \begin{tabular}{c}
    $\vf{A}{\Delta \cup \{\neg \neg B\},\Theta}$\\ \hline
    $\vf{A}{\Delta \cup \{B\},\Theta}$
   \end{tabular}
\end{tabular}
\bigskip

We need two more rules to obtain a complete system. The derivable rule EM0 and the
permissible rule IC greatly simplify the proof procedure.
\bigskip

\noindent
\begin{tabular}{@{}llll}
  Trans
 & \begin{tabular}{c}
    $\vf{A}{\Delta\cup\{B\},\Theta}$ \\
    $\vf{B}{\Delta',\Theta'}$ \\ \hline
    $\vf{A}{\Delta \cup \Delta',\Theta\cup\Theta'}$
   \end{tabular} \hspace{27mm}
 & EM
 & \begin{tabular}{c}
      $\vf{A}{\Delta \cup \{B\},\Theta}$\\
      $\vf{A}{\Delta' \cup \{\neg B\},\Theta'}$\\ \hline
      $\vf{A}{\Delta \cup \Delta',\Theta\cup\Theta'}$
     \end{tabular}
 \medskip \\
   EM0
 & \begin{tabular}{c}
      $\vf{A}{\Delta \cup \{\neg A\},\Theta}$\\ \hline
      $\vf{A}{\Delta,\Theta}$
   \end{tabular}
 & IC
 & \begin{tabular}{c}
    $\vf{\Dab(\Lambda \cup \Lambda')}{\Delta, \Theta\cup \Lambda'}$ \\ \hline
    $\vf{\Dab(\Lambda \cup \Lambda')}{\Delta, \Theta}$
   \end{tabular}
\end{tabular}
\bigskip

Each phase of the proof starts by applying a goal rule. All further steps proceed in
view of D-conditions of unmarked lines, or in view of A-conditions of unmarked
lines---see the procedure below. Premises are introduced and formulas analysed iff an
element of a D-condition is a \emph{positive part} of the added formula.

That $A$ \emph{is a positive part of} $B$ is defined as follows:%
\smallskip

\begin{rlist}{(viii)}
 \item[(i)]
  $A$ is a positive part of each of the following: $A$, $A\wedge B$, $B\wedge A$,
  $A\vee B$, $B\vee A$, $B\supset A$, $A\equiv B$, and $B\equiv A$;
 \item[(ii)]
  $A$ is a negative part of $\neg A$, $\krul A$, $A\supset B$, $A\equiv B$, and $B\equiv A$;
 \item[(iii)]
  if $A$ is a negative part of $B$, then $\neg A$ and $\krul A$ are positive parts of $B$.
 \item[(iv)]
  if $A$ is a positive part of $B$ and $B$ is a positive part of $C$, then $A$ is a
  positive part of $C$;
 \item[(v)]
  if $A$ is a positive part of $B$ and $B$ is a negative part of $C$, then $A$ is a
  negative part of $C$;
 \item[(vi)]
  if $A$ is a negative part of $B$ and $B$ is a positive part of $C$, then $A$ is a
  negative part of $C$;
 \item[(vii)]
  if $A$ is a negative part of $B$ and $B$ is a negative part of $C$, then $A$ is a
  positive part of $C$.
\end{rlist}
\smallskip

The efficiency of phase 3 and phase 1A is increased by defining, for those phases, $A$
as a positive part of $\neg\krul A$ and by dropping ``$\krul A$'' from clauses (ii) and
(iii).

A-marking (marking in view of the A-conditions, providing from the adaptive character of
the logic) is taken over by the procedure below. D-marking (marking in view of
D-conditions) is governed by the following definition.

\begin{definition}
Where $A_{\Delta,\Theta}$ occurs in the proof at line $i$,
 line $i$ is D-marked
 iff
 one of the following conditions is fulfilled:
 \begin{enumerate}
 \item \label{een}
   line $i$ is not an application of a goal rule and $A \in \Delta$,
 \item \label{twee}
   line $i$ is not an application of a goal rule
   and, for some $\Delta' \subset \Delta$ and $\Theta' \subseteq \Theta$,
   $A_{\Delta',\Theta'}$ occurs in the proof,
 \item
   no application of EFQ occurs in the proof and
   $B, \neg B \in \Delta$ for some $B$,
 \item
   no application of EFQ occurs in the proof and,
   for some $B \in \Delta$,
   $\neg B_{\emptyset,\emptyset}$ occurs in the proof at an unmarked line.
 \end{enumerate}
\end{definition}

If \ref{een} is the case, the condition is circular; if \ref{twee} is the case, some
(set theoretically) weaker condition is sufficient to obtain $A$. In the other two
cases, line $i$ indicates a search path that can only be successful if the premises are
$\neg$-inconsistent. Although it is not necessary to mark such search paths, it turns
out more efficient to postpone them to phase 1B.

I shall first present a rough outline of the procedure and next shall offer some
comments on fine tuning.

\paragraph{The procedure.}
The proof procedure for $\Gamma \vdash_\sys{ACLuN1} G$ consists of three phases---I
shall disregard infinite $\Gamma$. The procedure starts in phase 1, may move to phases 2
and 3, and returns to phase 1. During phases 2 and 3, a line may be A-marked (marked in
view of its A-condition). A phase stops if no lines can be added in view of conditions
introduced during that phase.

\paragraph{\textit{Phase 1.}}
Aim: to derive $G_{\emptyset, \Theta}$ for some $\Theta$. There are three cases:
\begin{rlist}{(0.0)}
 \item[(1.1)]
  $G_{\emptyset, \emptyset}$ is derived. Then $\Gamma \vdash_\sys{ACLuN1} G$.
 \item[(1.2)]
  $G_{\emptyset, \Theta}$ is derived, say at line $i$.
  The procedure moves to phase 2 and later returns to phase 1. There are two cases:
  \begin{rlist}{(1.0.0)}
   \item[(1.2.1)]
    line $i$ is not A-marked: $\Gamma \vdash_\sys{ACLuN1} G$.
   \item[(1.2.2)]
    line $i$ is A-marked: go on
    (aim: derive $G_{\emptyset, \Theta'}$ for some $\Theta' \nsupseteq \Theta$).
  \end{rlist}
 \item[(1.3)]
  The procedure stops and
  $G_{\emptyset, \Theta}$ is not derived on an unmarked line for any $\Theta$.
  Then $\Gamma \not\vdash_\sys{ACLuN1} G$.
\end{rlist}

\paragraph{\textit{Phase 2.}}
$G_{\emptyset, \Theta}$ was derived in phase 1 for some $\Theta$, say at line $i$. Phase
2 starts by applying A-Goal in order to add $\Dab(\Theta)_{\{\Dab(\Theta)\},
\emptyset}$. Aim: to derive $\Dab(\Theta)_{\emptyset, \Lambda}$ for some $\Lambda$
($\subseteq \Omega$). There are three cases:
\begin{rlist}{(0.0)}
 \item[(2.1)]
  $\Dab(\Theta)_{\emptyset, \emptyset}$ is derived:
  line $i$ is A-marked; the procedure returns to phase 1.
 \item[(2.2)]
  $\Dab(\Theta)_{\emptyset, \Lambda}$ is derived for some $\Lambda$, say at line $j$.
  The procedure moves to phase 3 and later returns to phase 2. There are two cases:
  \begin{rlist}{(1.0.0)}
   \item[(2.2.1)]
    line $j$ is A-marked: go on
    (aim: derive $\Dab(\Theta)_{\emptyset, \Lambda'}$ for some $\Lambda' \nsupseteq \Lambda$).
   \item[(2.2.2)]
    line $j$ is not A-marked: line $i$ is A-marked; the procedure returns to phase 1.
  \end{rlist}
 \item[(2.3)]
  $\Dab(\Theta)_{\emptyset, \Lambda}$ is not derived for any $\Lambda$ when phase 2 stops.
  Line $i$ is not A-marked and the procedure returns to phase 1.
\end{rlist}

\paragraph{\textit{Phase 3.}}
$G_{\emptyset, \Theta}$ was derived in phase 1 for some $\Theta$, say at line $i$, and
$\Dab(\Theta)_{\emptyset, \Lambda}$ was derived in phase 2 for some $\Lambda$, say at
line $j$. Phase 3 starts by applying X-Goal in order to add
$\Dab(\Lambda)_{\{\Dab(\Lambda)\}, \emptyset}$. Aim: to derive
$\Dab(\Lambda)_{\emptyset, \emptyset}$---all lines added in phase 3 should have the
A-condition $\emptyset$. There are two cases:
\begin{rlist}{(0.0)}
 \item[(3.1)]
  $\Dab(\Lambda)_{\emptyset, \emptyset}$ is derived:
  line $j$ is A-marked; the procedure returns to phase 2.
 \item[(3.2)]
  Phase 3 stops without $\Dab(\Lambda)_{\emptyset, \emptyset}$ being derived:
  line $j$ is not A-marked; the procedure returns to phase 2.
\end{rlist}

\paragraph{Some fine tuning.}
I shall start with some comments that concern the procedure itself, and next offer some
comments that pertain to the efficiency of the proofs.

The order in which one tries to apply rules is spelled out in \cite{metD:vd1}. The idea
is: first apply rules in order to obtain the goal of the current phase in a strictly
goal directed way, viz.\ by a sequence of applications of formula analysing rules,
condition analysing rules, Trans, EM0, and IC. Next, one tries to obtain the goal by
combining the former rules with applications of EM and Trans.

EFQ is never applied in phase 2 or 3. EFQ is only useful if the premises are
inconsistent. This is justified by the following consideration. EFQ can only be
successfully applied in a proof for $\Gamma \vdash G$ if $\Gamma$ is
$\neg$-inconsistent. In that case, $G_{\emptyset,\emptyset}$ is derivable from the
premises and will be derived in phase 1B. Deriving any \Dab-formula from $\Gamma$ by
applying EFQ (possibly combined with other rules) is a useless detour.

Moreover, EFQ is only applied in phase 1 at points where no other rule can be applied
and, from that point on---that is in subphase 1B---one adds only lines with an empty
A-condition to the proof, and hence never moves on to phase 2. The reason for this is
obvious: if the main goal can only be obtained by EFQ, then it is derivable by the lower
limit logic, viz.\ \sys{CLuN}, and hence there is no point in deriving it on some
A-condition.

I now describe an apparently rather efficient way of proceeding; it is nearly identical
for the three phases. Let me start with some general instructions. First, one never
applies a formula analysing rule on a formula that does not have a premise in its path.
Such steps are provably complications only. Moreover, no line is added to the proof if
it would at once be marked.

At each point after line 1 has been written, one first tries to apply EM0, EM and Trans
provided this leads to a line being marked.

If this fails, one proceeds in a strictly goal directed way. More particularly, one acts
in view of the first formula in the last unmarked condition (of the current phase). If
this formula cannot be obtained from the premises, then obtaining the other members of
the same condition is useless anyway. If no step is possible in view of the first
formula in the last unmarked condition of the current phase---this means that this
formula is a dead end---one acts in view of the first formula in the next-to-last
unmarked condition of the current phase, and so on.

If it is possible to act in view of the first formula of an unmarked condition of the
current phase, one applies the rules in the following order---remember what was said
about positive parts. First one tries to apply a formula analysing rule on a formula
that occurs on an unmarked line. Next, one tries to introduce a premise. Finally one
applies a condition analysing rule (to the formula in view of which one proceeds).

If the goal of the current phase cannot be obtained by strictly goal directed moves, one
also applies Trans in order to obtain the goal on all possible (unmarked) conditions,
and next one applies EM to unmarked lines that have the current goal as their second
element.%
 \footnote{Apparently this application of EM is useless, but our proof in
 \cite{metD:vd1} that the procedure is complete relies on it.}
One returns to strictly goal directed moves as soon as possible.

Only if all this fails, one applies EFQ in phase 1 and, as said before, from there on
only adds lines with an empty A-condition.

\paragraph{Some comments on the metatheory.}
The procedure is provably an algorithm for $\Gamma \vdash_\sys{ACLuN1} A$. In
\cite{metD:vd1} this is proved for \sys{CL}. That proof can easily be transformed to
show that the rules from the present section are sound and complete with respect to
\sys{CLuN} in the following sense (for finite $\Gamma$):
\begin{rlist}{(0)}
 \item[(1)]
  If $\Gamma \vdash_\sys{CLuN} G$,
  then
  $G_{\emptyset,\emptyset}$ is derived in the dynamic proof for $\Gamma \vdash_\sys{CLuN} G$.\\
  If $\Gamma \not\vdash_\sys{CLuN} G$,
  then
  the dynamic proof for $\Gamma \vdash_\sys{CLuN} G$ stops.
 \item[(2)]
  $A_{\emptyset,\Theta}$ is derivable in the dynamic proof for $\Gamma \vdash_\sys{CLuN} G$
  iff
  $\Gamma \vdash_\sys{CLuN} A\vee\Dab(\Theta)$\,.
\end{rlist}

Given this, it is easily seen that the procedure is sound and complete with respect to
\sys{ACLuN1}.

An essential point concerns phase 2. Suppose that $G_{\emptyset, \Theta}$ is derived at
line $i$ for some $\Theta$, and that $\Dab(\Theta)_{\emptyset, \Lambda}$ is derived for
some $\Lambda$ at line $j$. It follows that $\Gamma \vdash_\sys{ACLuN1} \Dab(\Theta \cup
\Lambda)$. If $\Dab(\Lambda)_{\emptyset, \emptyset}$ is derived in phase 3, then $\Gamma
\vdash_\sys{ACLuN1} \Dab(\Lambda)$, and hence $\Dab(\Theta \cup \Lambda)$ is not a
minimal \Dab-consequence of $\Gamma$. So, $\Theta \cap U(\Gamma) = \emptyset$ iff the
following holds for all $\Lambda$: if $\Gamma \vdash_\sys{ACLuN1} \Dab(\Theta \cup
\Lambda)$, then $\Gamma \vdash_\sys{ACLuN1} \Dab(\Lambda)$. This condition comes to: if
$\Dab(\Theta)_{\emptyset, \Lambda}$ is derivable, then $\Dab(\Lambda)_{\emptyset,
\emptyset}$ is derivable.

Precisely this is checked in phase 2: the procedure returns to phase 1 with line $i$ not
A-marked iff it holds for all $\Lambda$ that $\Dab(\Lambda)_{\emptyset, \emptyset}$ is
derivable whenever $\Dab(\Theta)_{\emptyset, \Lambda}$ is derivable. So, if the
procedure returns to phase 1 with line $i$ not A-marked, then $\Theta \cap U(\Gamma) =
\emptyset$ and hence $G$ is finally derived at line $i$.

It is equally easy to see that line $i$ is marked just in case $\Theta \cap U(\Gamma)
\neq \emptyset$. If, for some $\Lambda$, $\Dab(\Theta)_{\emptyset, \Lambda}$ is
derivable whereas $\Dab(\Lambda)_{\emptyset, \emptyset}$ is not derivable, then $\Gamma
\vdash_\sys{ACLuN1} \Dab(\Theta \cup \Lambda)$ whereas $\Gamma \not\vdash_\sys{ACLuN1}
\Dab(\Lambda)$. It follows that $\Theta \cap U(\Gamma) \neq \emptyset$.%
 \footnote{Indeed, if $\Gamma \vdash_\sys{ACLuN1} \Dab(\Theta \cup \Lambda)$ and
 $\Gamma \not\vdash_\sys{ACLuN1} \Dab(\Lambda)$,
 there is a non-empty $\Theta' \subseteq \Theta$ and a (possibly
 empty) $\Lambda' \subseteq \Lambda$ such that $\Dab(\Theta' \cup \Lambda')$ is a
 minimal \Dab-consequence of $\Gamma$.}

\paragraph{Some examples.}
Let us start with two simple examples. Consider first a goal directed proof for $\krul
p\vee r, p\wedge \krul q, q \vdash_\sys{ACLuN1} r$\,:

\begin{tabbing}
00Mm \= $(\cdic{p})\vee$m \= 00, 00M \= Premmm \= $\{\cdic{p}\}$m \= \kill
 1 \> $r$ \>  \> Goal \> $\{r\}$ \> $\emptyset$\\
 2 \> $\krul p\vee r$ \>  \> Prem \> $\emptyset$ \> $\emptyset$\\
 3 \> $r$ \> 2 \> $\vee$E \> $\{\neg\krul p\}$ \> $\emptyset$\\
 4 \> $r$ \> 3 \> C$\neg\krul$E \> $\{p\}$ \> $\{\cdic{p}\}$\\
 5 \> $p\wedge \krul q$ \>  \> Prem \> $\emptyset$ \> $\emptyset$\\
 6 \> $p$ \> 5 \> $\wedge$E \> $\emptyset$ \> $\emptyset$\\
 7 \> $r$ \> 4, 6 \> Trans \> $\emptyset$ \> $\{\cdic{p}\}$\\
 8 \> $\cdic{p}$ \>  \>A-Goal \> $\{\cdic{p}\}$ \> $\emptyset$\\
 9 \> $\cdic{p}$ \> 8 \> C$\wedge$E \> $\{\cdic{p}\}$ \> $\emptyset$\\
 10\> $\cdic{p}$ \> 6, 9 \> Trans \> $\{\krul p\}$ \> $\emptyset$\\
 11\> $\krul p$ \> 2 \> $\vee$E \> $\{\neg r\}$ \> $\emptyset$\\
 12\> $\cdic{p}$ \> 10 \> C$\krul$E \> $\{\neg p\}$ \> $\emptyset$\\
 13\> $\neg p$ \> 11 \> $\krul$E \> $\{\neg r\}$ \> $\{\cdic{p}\}$\\
 14\> $\cdic{p}$ \> 12, 13 \> Trans \> $\{\neg r\}$ \> $\{\cdic{p}\}$\\
 15\> $\cdic{p}$ \> 14 \> IC \> $\{\neg r\}$ \> $\emptyset$
\end{tabbing}
The proof is successful: at line 7 $r$ is derived on the empty D-condition and on the
A-condition $\{\cdic{p}\}$, and in phase 2 $\cdic{p}$ turns out not to be derivable on
any A-condition. The situation is similar whenever $G_{\emptyset,\Theta}$ is derivable
and $\Dab({\Theta})_{\emptyset,\Lambda}$ is not derivable for any $\Lambda$. Remark that
this always obtains if the premise set is $\krul$-consistent.

Next, consider the goal directed proof for $\krul p, p\vee q, p \vdash_\sys{ACLuN1}
q$\,:

\begin{tabbing}
00Mm \= $(\cdic{p})\vee$m \= 00, 00M \= Premmm \= $\{\cdic{p}\}$m \= \kill
 1 \> $q$ \>  \> Goal \> $\{q\}$ \> $\emptyset$\\
 2 \> $p\vee q$ \>  \> Prem \> $\emptyset$ \> $\emptyset$\\
 3 \> $q$ \> 2 \> $\vee$E \> $\{\neg p\}$ \> $\emptyset$\\
 4 \> $\krul p$ \>  \> Prem \> $\emptyset$ \> $\emptyset$\\
 5 \> $\neg p$ \> 4 \> $\krul$E \> $\emptyset$ \> $\{\cdic{p}\}$\\
 6 \> $q$ \> 3, 5 \> Trans \> $\emptyset$ \> $\{\cdic{p}\}$\\
 7 \> $\cdic{p}$ \> \> A-Goal \> $\{\cdic{p}\}$ \> $\emptyset$\\
 8 \> $\cdic{p}$ \> 7 \> C$\wedge$E \> $\{p, \krul p\}$ \> $\emptyset$\\
 9 \> $\cdic{p}$ \> 4, 8 \> Trans \> $\{p\}$ \> $\emptyset$\\
 10\> $p$ \> 2 \> $\vee$E \> $\{\neg q\}$ \> $\emptyset$\\
 11\> $p$ \>  \> Prem \> $\emptyset$ \> $\emptyset$\\
 12\> $\cdic{p}$ \> 9, 11 \> Trans \> $\emptyset$ \> $\emptyset$\\
 13\> $q$ \>   \> EFQ \> $\{\neg (p\vee q)\}$ \> $\emptyset$\\
 14\> $q$ \>   \> EFQ \> $\{\neg\krul p\}$ \> $\emptyset$
\end{tabbing}
After $q_{\{\cdic{p}\},\emptyset}$ is derived at line 6,
$\cdic{p}_{\emptyset,\emptyset}$ turns out to be derivable (line 12). The procedure then
sets out to derive $q$ in a different way, which fails. Neither variant of C$\vee$E is
applied to the condition of line 13 because the resulting line would at once be marked.
C$\neg\krul$E is not applied to the condition of line 14 because doing so would
introduce a non-empty A-condition.

Finally, let us consider the goal directed proof for $p, \krul p\vee s, r\supset t,
\krul p\vee q, \krul q \vdash_\sys{ACLuN1} s$\,:

\begin{tabbing}
00Mm \= $(\cdic{p})\vee$m \= 00, 00M \= Premmm \= $\{\cdic{p}\}$m \= \kill
 1 \> $s$ \>  \> Goal \> $\{s\}$ \> $\emptyset$\\
 2 \> $\krul p\vee s$ \>  \> Prem \> $\emptyset$ \> $\emptyset$ \\
 3 \> $s$ \> 2 \> $\vee$E \> $\{\neg\krul p\}$ \> $\emptyset$ \\
 4 \> $s$ \> 3 \> C$\neg\krul$E \> $\{p\}$ \> $\{\cdic{p}\}$ \\
 5 \> $p$ \>  \> Prem \> $\emptyset$ \> $\emptyset$ \\
 6 \> $s$ \> 4, 5 \> Trans \> $\emptyset$ \> $\{\cdic{p}\}$ \\
 7 \> $\cdic{p}$ \>  \> A-Goal \> $\{\cdic{p}\}$ \> $\emptyset$ \\
 8 \> $\cdic{p}$ \> 7 \> C$\wedge$E \> $\{p, \krul p\}$ \> $\emptyset$ \\
 9 \> $\cdic{p}$ \> 8, 5 \> Trans \> $\{\krul p\}$ \> $\emptyset$ \\
 10\> $\krul p$ \> 2 \> $\vee$E \> $\{\neg s\}$ \> $\emptyset$ \\
 11\> $\krul p\vee q$ \>  \> Prem \> $\emptyset$ \> $\emptyset$ \\
 12\> $\krul p$ \> 11 \> $\vee$E \> $\{\neg q\}$ \> $\emptyset$ \\
 13\> $\krul q$ \>  \> Prem \> $\emptyset$ \> $\emptyset$ \\
 14\> $\neg q$ \> 13 \> $\krul$E \> $\emptyset$ \> $\{\cdic{q}\}$ \\
 15\> $\krul p$ \> 12, 14 \> Trans \> $\emptyset$ \> $\{\cdic{q}\}$ \\
 16\> $\cdic{p}$ \> 9, 15 \> Trans \> $\emptyset$ \> $\{\cdic{q}\}$ \\
 17\> $\cdic{q}$ \>  \> X-Goal \> $\{\cdic{q}\}$ \> $\emptyset$ \\
 18\> $\cdic{q}$ \> 17 \> C$\wedge$E \> $\{q, \krul q\}$ \> $\emptyset$ \\
 19\> $\cdic{q}$ \> 13, 18 \> Trans \> $\{q\}$ \> $\emptyset$ \\
 20\> $q$ \> 11 \> $\vee$E \> $\{\neg\krul p\}$ \> $\emptyset$
\end{tabbing}
Here phase 3 stops, $\neg\krul p$ not being \sys{CLuN}-derivable from the premises. Line
16 is not A-marked and the procedure returns to phase 2; there line 6 is A-marked and
the procedure returns to phase 1. The procedure there aims at deriving $s_{\emptyset,
\Theta}$ in phase 1 for some $\Theta \nsupseteq \{\cdic{p}\}$, which fails.

It is instructive to study the procedure and consider the different states in which it
may stop in phase 1.

A computer programme that implements the procedure is available---the above proofs are
produced by it. The programme will be used for presenting further examples during the
lecture and will be on the internet before this paper appears---\texttt{\small
http://logica.rug.ac.be/dirk/}. The data file that goes with the programme contains a
set of instructive example exercises.

\section{In Conclusion}

The `defeasible' conditions that occur in dynamic proofs of adaptive logics suggested a
kind of dynamic proofs with `prospective' conditions. This led to a specific form of
goal directed proofs. Later, these goal directed proofs turned out to provide a proof
procedure that forms an algorithm for final derivability at the propositional level. As
remarked in Section \ref{aim}, the central interest of the procedure is that it provides
a criterion at the predicative level if it stops.

The dynamic proofs explicate actual reasoning. The goal directed proofs do not, but
there is an algorithm for turning them into dynamic proofs (by reordering and replacing
lines). So, after finding out that some formula is derivable at a stage from the
premises, one may switch to the goal directed format in order to find out whether the
formula is finally derivable. If a decision is reached, one may transform the result to
a regular dynamic proof, if desired. After this, the proof may proceed and, if a further
interesting formula is derived at a stage, one may again switch to the goal directed
format to settle its final derivability.

This seems the right place to insert a comment on the original application context
mentioned in Section \ref{aclun1}. It was proved in \cite{ial} that $\Dab(\Delta\cup \{
A\})$ is not a minimal \Dab-consequence of $\Gamma$ unless $\krul A$ is a subformula of
some member of $\Gamma$.%
 \footnote{At the predicative level, $A$ may be an open formula in which case the
 corresponding abnormality is the existential closure of $\cdic{A}$. The criterion in
 the text has then to be modified to: $\krul B$ is a subformula of a member of $\Gamma$,
 where $B$ is obtained by relettering the individual variables in $A$.}
In view of this, the goal directed proofs provide a means to locate all minimal
\Dab-consequences of finite premise sets.

Given the present standard characterization (from \cite{gcal}) of flat adaptive logics,
some minimal changes to the aforementioned rules will result in a goal directed
procedure for any other adaptive logic. Basically, one replaces the rules that pertain
to the abnormalities---in the case of \sys{ACLuN1}, the rules containing the
paraconsistent negation $\krul$.

While these replacements are straightforward, further research is required for the
predicative level. Devising sensible rules is unproblematic---the relevant research was
finished. However, more work is needed to improve the efficiency of the procedure and to
avoid infinite loops whenever possible. It is easily seen that known techniques from
tableau methods and resolution methods may easily be transposed to the goal directed
proofs.%
 \footnote{Unpublished papers by members of our research group are available
 from the internet address \texttt{\small http://logica.rug.ac.be/centrum/writings/}.}

\end{document}